 \documentclass[12pt,preprint]{aastex}

\usepackage{graphicx}				         
\usepackage{amsmath}
\usepackage{enumitem}			          	
\usepackage{amssymb}
\usepackage{pdfpages}

\newcommand {\HST}  {{\it HST}}        
\newcommand {\FUSE}  {{\it FUSE}}   

\newcommand {\Lya}    {Ly$\alpha$}   

\newcommand {\OI}      {\ion{O}{1}}   
\newcommand {\OII}     {\ion{O}{2}}   
\newcommand {\OIII}    {\ion{O}{3}}   
\newcommand {\OIV}    {\ion{O}{4}}   
\newcommand {\OVI}    {\ion{O}{6}}   

\newcommand {\SiII}     {\ion{Si}{2}}   
\newcommand {\SiIII}    {\ion{Si}{3}}   
\newcommand {\SiIV}     {\ion{Si}{4}}   

\newcommand {\CIII}     {\ion{C}{3}}   
\newcommand {\CIV}     {\ion{C}{4}}   
\newcommand {\CV}      {\ion{C}{4}}   

\newcommand {\NIV}      {\ion{N}{4}}   
\newcommand {\NV}      {\ion{N}{5}}    

\newcommand {\NeVIII}  {\ion{Ne}{8}}   

\newcommand {\MgII}     {\ion{Mg}{2}}   

\newcommand {\etal} {et~al.}

\newcommand {\kms}    {km~s$^{-1}$}

\newcommand {\HI}     {\ion{H}{1}}      
\newcommand {\HII}    {\ion{H}{2}}      
\newcommand {\HeI}    {\ion{He}{1}}   
\newcommand {\HeII}   {\ion{He}{2}}   
\newcommand {\HeIII}  {\ion{He}{3}}   

\begin{document}
\title{Softening of the Metagalactic Ionizing Background from  \\
    Internal \HeII\ Absorption in Quasars}
\author{J. Michael Shull \& Charles W. Danforth}
\affil {CASA, Department of Astrophysical \& Planetary Sciences. \\
University of Colorado, Boulder, CO 80309}

\email{michael.shull@colorado.edu, danforth@colorado.edu}



\begin{abstract}

Quasars and other active galactic nuclei (AGN) are significant contributors to the metagalactic 
ionizing background at redshifts $z < 3$.  Recent \HST/COS composite spectra of AGN find a 
harder flux distribution in the Lyman continuum, $F_{\nu} \propto \nu^{-\alpha_s}$ 
($\alpha_s = 1.41\pm0.15$) compared to previous studies.  This index appears to be inconsistent 
with observed \HeII/\HI\ absorption ratios ($\eta$) in the \Lya\ forest.   We explore the effects of 
internal AGN absorption in the \HeII\ (4-ryd) continuum using an analytic source-function model 
of the ionizing background in which the emissivity ($j_{\nu}$) arises from quasars, reprocessed 
by the opacity ($\kappa_{\nu}$) of the intervening \Lya\ forest and distinct AGN escape fractions 
$f_{\rm esc}^{(\rm HI)}$ and $f_{\rm esc}^{(\rm HeII)}$ at 1~ryd and 4~ryd, respectively.   We
also examine \HI\ and \HeII\ photoelectric heating from the reprocessed background, whose
spectral index ($\alpha_b > \alpha_s)$ depends on $\alpha_s$, the \HI\ column density slope 
$\beta$, and the ratio $R = f_{\rm esc}^{\rm (HI)} /  f_{\rm esc}^{(\rm HeII)}$.  
We compare the model to \Lya\ absorption lines of \HeII\ and \HI\ toward the quasar HE~2347-4342.
Internal AGN absorption with $f_{\rm esc}^{\rm (HeII)} \approx 0.6-0.8$ but 
$f_{\rm esc}^{\rm (HI)} \approx 1$ would increase the index by $\Delta \alpha_b  \approx 0.3-1.0$,
corresponding to $\eta = 60-200$ for $\beta \approx 1.5-1.6$, in agreement with HST/COS 
observations at $z \approx 2.5-2.9$.   The observed range of ratios, $\eta < 200$, constrains 
$\alpha_b  < 3.4$ and $f_{\rm esc}^{(\rm HeII)}  > 0.4$.  Individual AGN with softer spectra,
$\alpha_s > 1.7$, and more internal \HeII\ absorption could produce a few absorbers with 
$\eta > 300$, in proximity to AGN transverse to the sight line.  

\vspace{3cm}

\end{abstract}

\section{Introduction}

Quasars and other active galactic nuclei (AGN) are considered to be significant contributors to the 
metagalactic ionizing background radiation (Haardt \& Madau 2012), affecting the physical state of both 
the intergalactic medium (IGM) and circumgalactic medium (CGM).   The dominant sources of ionizing 
photons are massive stars within star-forming galaxies and black hole accretion disks in AGN, but their
relative contributions remain uncertain (Madau \& Haardt 2015;  Finkelstein \etal\  2019;  Puchwein \etal\ 
2019).  Although the space density of quasars is much smaller than star-forming galaxies (Richards \etal\ 
2006; Madau \& Dickinson 2014) a larger fraction of ionizing radiation escapes from luminous AGN.   
Far-ultraviolet spectra from the Cosmic Origins Spectrograph (COS; Green \etal\ 2012) on the 
{\it Hubble Space Telescope} (\HST) have measured the Lyman continuum of low-redshift ($z \leq 1.5$) 
quasars down to rest-frame wavelengths of 350~\AA.   Composite spectra from ensembles of Type-1 AGN 
(Stevans \etal\ 2014; Tilton \etal\ 2016) along clear sight lines through their emission-line regions and 
host galaxies show no photoionization edges at 912~\AA\ (\HI) or 504~\AA\ (\HeI) suggesting an escape 
fraction $f_{\rm esc}^{(\rm HI)} \approx 1$ for  the \HI\ ionizing continuum at 1~ryd. 
Internal AGN absorption in the higher-opacity \HeII\ continuum is likely to exist, although there is no direct 
information on the \HeII\ edge (228~\AA) owing to strong absorption in the \HI\ \Lya\ forest at $z > 2$.
Much smaller \HI\ escape fractions or upper limits of a few percent are observed for star-forming galaxies 
(Shapley \etal\ 2016; Vanzella \etal\ 2018;  Izotov \etal\ 2018) and \Lya\ emitters (Bian \& Fan 2020).  

Photoionizing radiation is likely responsible for two major transitions of the IGM during the reionization
epochs of \HI\  at redshift $z \approx 7$ and \HeII\  at $z \approx 3$.  Obtaining a deeper understanding of 
the processes governing these transitions requires knowledge of the production rates of ionizing photons 
from high-redshift  galaxies and AGN, as well as the range of escape fractions and their dependence on 
galaxy mass, metallicity, and stellar properties.   Quasars generally have hard spectra in the 
\HeII\ continuum, whereas hot  O-type stars have \HeII\ photoionization edges at 228~\AA.  The ionizing 
spectra of hot stars and AGN should be distinguishable by their intrinsic spectra and their effects on the 
state of the surrounding gas.  Commonly used diagnostics of the EUV background include abundances 
of metal ions in the IGM and CGM as well as the \HeII/\HI\ ratio in the \Lya\ forest  (Kriss \etal\ 2001; 
Shull \etal\ 2010) and Lyman Limit Systems (Graziani \etal\ 2019).  

There are important cosmological reasons for identifying the populations that produce \HeII\ reionization.
Quasar absorption lines in the \Lya\ transitions of  \HI\ (1215.67~\AA) and \HeII\ (303.78~\AA) provide 
evidence for a ``cosmic web" of  gaseous structures in the IGM.  During the \HeII\ reionization epoch, the 
\HeIII\ ionization fronts begin to overlap and the IGM gradually becomes transparent in the 4 ryd \HeII\ continuum.   
The luminosity function of AGN and its evolution in redshift define the duration of the transition, leaving a 
thermal imprint on the gas (Puchwein \etal\ 2015), equation of state (Hui \& Gnedin 1997), and abundances
of metal-ions in IGM and CGM (Madau \& Haardt 2009;  Boksenberg \& Sargent 2015).  As AGN populations
evolve, the spectral characteristics of ionizing sources will influence the structure of the \Lya\ absorption 
troughs and flux transmission windows in the spectra of high-redshift AGN (Gunn \& Peterson 1965).  
Finally, \HeII-ionizing QSOs transverse to the sight line can be used to constrain QSO lifetimes (Jakobsen 
\etal\ 2003; Syphers \& Shull 2014;  Schmidt \etal\ 2018) and the propagation of cosmological ionization fronts.   

In this paper, we develop an analytic model of the ionizing background due to AGN with internal absorption 
in their continuum and emission-line regions.  This model employs the radiative source function in the diffuse 
IGM, where the intrinsic AGN spectral emissivity is modified by escape fractions  $f_{\rm esc}^{(\rm HeII)} < 1$
and $f_{\rm esc}^{\rm (HI)} \approx 1$.  These assumptions are consistent with observations of AGN in their 
rest-frame EUV at 912~\AA\ (\HI) and 504~\AA\ (\HeI) and inferred at 228~\AA\ (\HeII) from variations in the 
extent of \HeII\ proximity zones around QSOs.  Internal absorption in the \HeII\ continuum will produce a 
softer metagalactic background, consistent with observed \HeII/\HI\ ratios (denoted $\eta$) in the \Lya-forest.  
Throughout this paper, we assume that the IGM opacity comes from \HI\ absorbers with a power-law distribution 
in \HI\ column density, $f(N) \propto N^{-\beta}$, and that the ionizing background at $z < 3$ is dominated by 
AGN with a mean flux distribution in frequency, $F_{\nu} \propto \nu^{-\alpha_s}$.  
Here,  $\alpha_s$ is the intrinsic spectral index (positive by convention) characterizing the AGN ionizing 
continua of \HI\ and \HeII.   After absorption and reprocessing, the metagalactic ionizing background is 
characterized by an effective index $\alpha_b > \alpha_s$, defined by the ratio of specific intensities at
4 ryd and 1 ryd,  $J_{\rm HeII} / J_{\rm HI} \equiv 4^{-\alpha_b}$. As discussed in Section 3, this effective 
index can be related to the \HeII/\HI\ ratio by the formula $\eta =1.77 \times 4^{\alpha_b}$, where $\alpha_b$
depends on $\alpha_s$ and $\beta$.   The \HeII\ ion is typically much more abundant in the IGM than \HI\ 
(Miralda-Escud\'e \etal\ 1996; Fardal \etal\ 1998) even though He/H elemental abundances would favor 
hydrogen.  The large observed ratios of column densities, $N_{\rm HeII}/N_{\rm HI} \approx 50-200$, arise 
because \HeI\ is harder to ionize than \HI\ and  \HeIII\ recombines faster than \HII.  

Our new study is motivated by two recent  \HST/COS observations:  
(1) the composite ionizing spectrum of AGN is harder than previously assumed; and
(2) higher-resolution observations of the \Lya\ forest better define the range of $\eta$ ratios.  
To probe the \HeII\ post-reionization epoch, we obtained deep UV spectroscopic observations of two 
quasars,  HE\,2347-4342 and HS\,1700+6416, using both the G130M and G140L gratings of COS.  
These two AGN are the brightest known \HeII\ quasars and among the few percent of $z \sim 3$ quasars 
with sufficient far-UV flux to observe the \HeII\ \Lya\ forest in absorption. These were the only two quasars 
studied by \FUSE\ in \HeII\ absorption (Kriss \etal\ 2001;  Shull \etal\ 2004;  Zheng \etal\ 2004; 
Fechner \etal\ 2006).  These UV targets were  used by the COS GTO team and others to study the \HeII\ 
post-reionization epoch (Shull \etal\ 2010; Syphers \& Shull 2013; Worseck \etal\ 2016) and \HeII\ 
transverse-proximity effects in the ionization zones around QSOs (Schmidt \etal\ 2018).  

In Section 2, we describe our \HST/COS observations of HE~2347-4342, using the moderate-resolution
G130M grating in the super-blue (1222~\AA) wavelength setting, together with the lower-resolution 
G140L grating.  This allowed us to extend far-UV G130M coverage down to 1067~\AA\ and analyze 
\HeII/\HI\  ratios down to redshifts $z \approx 2.4-2.5$, well into the post-reionization epoch.  
Section 3 presents our analytic model of the reprocessing of the intrinsic AGN spectra by intervening 
IGM and by absorption internal to the AGN.  We introduce distinct escape fractions ($f_{\rm esc}$) for 
the \HI\ and \HeII\ continua and derive a formula for the effective spectral index $\alpha_b$ of the softened 
metagalactic radiation field and the corresponding \HeII/\HI\ ratios.  Adopting a mean index $\alpha_s = 1.4$, 
an $N_{\rm HI}$ distribution slope $\beta = 1.5\pm0.05$, and assuming no internal AGN absorption, we find 
$\alpha_b = 2.0-2.4$ and $\eta \approx 30-50$.   These ratios are considerably lower than those in the
observations, and additional spectral softening may be needed, depending on the slope of the \HI\ 
distribution between $15 < \log N_{\rm HI} < 17$.   We propose that \HeII-ionizing (4 ryd) photons are 
absorbed within some AGN, parameterized by escape fractions $f_{\rm esc}^{(\rm HeII)} \approx 0.6-0.9$ 
required for consistency with the $\eta$ distribution.  
Section 4 summarizes the implications of the modified radiation field for the \HeII\ reionization epoch 
and photoionization of metal ions in the IGM and CGM.  We suggest the need for more 
sophisticated models in which AGN are non-uniform  populations with a range of intrinsic spectral 
parameters ($\alpha_s$, $f_{\rm esc}$).  This would facilitate a study of their correlation with $\eta$ and 
influence on metal-ion abundance ratios such \CIV/\CIII, \SiIV/\SiIII, and \OVI/\NV.

\section{Observations}  

\subsection{AGN Composite Spectrum}

 The \HeII/\HI\ absorption ratio constrains the mean spectrum of quasars, after correcting for  absorption 
 and reprocessing by intervening gas.  Using the G130M/G160M gratings  on \HST/COS, 
Stevans \etal\ (2014) constructed a composite spectrum of 159 AGN ($z < 1.476$) in the rest-frame 
 EUV down to $\lambda_{\rm rest} \approx 475$~\AA.  The mean spectral index,  
 $\alpha_s = 1.41 \pm 0.15$, is harder than those in other \HST\  studies with lower-resolution instruments.  
 For example, Telfer \etal\ (2002) found $\alpha_s = 1.57 \pm 0.17$ for 39 radio-quiet QSOs observed 
 with the lower-resolution Faint Object Spectrograph (FOS) on \HST.   
 Lusso \etal\ (2015)  found $\alpha_s = 1.70 \pm 0.61$ for 53 blue QSOs at $z \approx 2.4$ using the 
 UVIS-G280 grism on the \HST\ Wide Field Camera~3 (2000--6000~\AA).  The \HST/COS composite 
 (Stevans \etal\ 2014) was based on high-S/N spectra with 20~\kms\  resolution, allowing us
 to resolve and mask out intervening \Lya-forest absorbers. 
 The high S/N allowed us to identify broad emission lines from the AGN (\NeVIII, \OIV, and other ions of 
 O and Ne) and characterize the underlying continuum.  We also identified Lyman Limit systems (LLS) 
 and partial LLS  (pLLS) and restored the true continuum. Their \HI\ column densities were determined 
 from the depth of the Lyman break and curve-of-growth analysis of Lyman series lines (Shull \etal\ 2017).
 
We believe the COS  composite index, $\alpha_s = 1.41\pm0.15$, is a more reliable measure of the 
EUV (1-2 ryd) spectrum of AGN than the lower-resolution studies with FOS and WFC3.   The quoted error 
bars on $\alpha_s$ ($\pm 0.15$ with COS, $\pm 0.17$ with FOS, $\pm 0.61$ with WFC3) allow for some 
overlapping uncertainty in the spectral index.   However, as we show in Section 3, the effective index 
$\alpha_b$ changes rapidly with IGM reprocessing, as the imean ntrinsic index hardens from 
$\alpha_s = 1.7$ to 1.4.  These changes produce significant differences in predicted \HeII/\HI\ ratios
 in models with $\alpha_s = 1.4$ (COS) versus $\alpha_s = 1.7$ (WFC3 prism).

\subsection{Previous Studies of \HeII }  

Our previous COS/GTO analysis of  HE\,2347-4342 (Shull \etal\ 2010) was based on G130M spectra
between 1135--1440~\AA\ at 20 \kms\ resolution.  Because of the central wavelength setting, these 
spectra did not cover \HeII\ \Lya-forest absorption in the post-reionization recovery at $z < 2.75$.   
The shorter UV wavelengths were only observed at low resolution ($\sim200$~\kms) with G140L on
\HST/COS.  Earlier  \FUSE\ observations (Kriss \etal\ 2001; Shull \etal\ 2004) had low throughput and 
uncertain background corrections.   The quasar HS\,1700+6416 was previously observed by 
COS/G140L (Syphers \& Shull 2014) but at low resolution insufficient to see individual absorbers.   
Our new \HST/COS data with G130M provide superior resolution of the \Lya\ forest in \HeII.  

In this paper, we describe properties of the \Lya\ forest toward HE\,2347-4342.  This sight line is 
considerably less complicated, compared to HS\,1700+6416, which has contamination from metal 
lines in several strong intervening systems in (Fechner \etal\ 2006).   The COS/G130M spectrum of 
HE\,2347-4342 showed strong \HeII\ ($\lambda 303.78$) absorption extending 1480~\kms\ redward 
of the estimated systemic redshift ($z_{\rm sys} \approx 2.885$) out to 1186.4~\AA\ ($z_a = 2.905$).  
Absorption at $z_a \geq z_{\rm QSO}$ is also seen in \HI\ and in associated metal-line absorbers 
(Fechner \etal\ 2004).   Previous studies of the weak (\OI+\SiII) $\lambda1302$ emission-line blend 
quoted systemic redshifts for HE\,2347-4342 of $z_{\rm sys} = 2.885 \pm 0.005$ (Reimers \etal\ 1997), 
$z_{\rm sys} = 2.885 \pm0.003$ (Dall'Aglio \etal\ 2008a), and $z_{\rm sys} = 2.886 \pm 0.003$ 
(Dall'Aglio \etal\ 2008b).   There is no obvious ``proximity effect" (flux transmission shortward of the 
\HeII\ edge) despite the large ionizing photon luminosity of the QSO in the \HI\ and \HeII\ continua.   
The observed strong associated absorption (1181.0--1186.4~\AA) and the lack of a QSO proximity 
effect prompted the suggestion (Shull \etal\ 2010) that the true QSO redshift might be slightly higher, 
$z_{\rm sys} = 2.904\pm0.002$, just  longward of the associated absorbers.  That hypothesis turned 
out to be incorrect.   Near-infrared spectra taken with the FIRE  (Folded-port InfraRed Echellette) 
spectrograph at the Magellan 6.5m telescope (R. Simcoe, 2015, private communication) measured  
$z_{\rm sys} = 2.886\pm0.001$ from the lines of [\OIII], H$\gamma$, and \MgII.  This result is consistent 
with previous redshift  estimates from the broad \OI\ $\lambda1302$ emission line, with improved errors.  
This more accurate systemic redshift is important for understanding the ionizing radiation field near 
the QSO, and opens the possibility that some \HeII-ionizing photons are being absorbed within the AGN.  
 
 Previous COS observations of HE~2347-4342 (Shull \etal\ 2010) found substantial variations in the
 \HeII/\HI\ optical-depth ratios ($\eta$) with a mean value $\langle \log \eta \rangle = 1.9$.  Because of 
 binning and line overlap, the $\eta$ values depend on spectral resolution:  $\eta =10-100$ for 
 $2.4 < z < 2.73$ with the low-resolution G140L grating, and $\eta = 5-500$ for $2.75 < z < 2.89$ with 
 the higher resolution G130M grating.  These $\eta$-variations were seen at small scales 
 ($\Delta z \approx 0.01$).  For HS~1700+6416, a similar range of ratios were found, with
 $\langle \log \eta \rangle = 1.77$ and median $\eta = 1.86-1.90$ from \FUSE\ data (Fechner \etal\ 2006) 
 and $\langle \eta \rangle = 73^{+10}_{-11}$ from \HST/COS data  taken with the lower resolution G140L
 grating (Syphers \& Shulll 2013).

 \subsection{New Observations of HE\,2347-4342 }
 
We obtained high-S/N data at moderate resolution ($\sim$30 \kms) with the COS/G130M grating covering 
1067--1363 \AA\ (see Table 1).   The \HeII\ data were obtained through \HST\ Guest Observer Program 13301
(PI Shull) using the super-blue (1222~\AA) setting and near-simultaneous spectra with the lower-resolution 
G140L grating. The HST/COS G130M spectrograph is ideal for high-S/N studies of individual far-UV absorbers, 
while G140L is used to define the AGN continuum longward of the \HeII\ edge.  
We chose the 1222~\AA\ setting, rather than 1096~\AA\  or 1055~\AA, because it provides higher resolving 
power ($R \approx 15,000$ rather than 6000) over the wavelengths (1067--1140~\AA) of  most interest for 
the \HeII\ post-reionization era.    This setting has higher throughput than the other two, and it produces the 
higher S/N needed to detect and resolve the \HeII\ \Lya\ absorbers at $z < 2.75$.  The G130M/1222~\AA\ setting
also allows sufficient anchoring continuum longward of the target \HeII\ \Lya\ break at 1186~\AA.
With these data (14 \HST\ orbits) we were able to resolve individual lines in the \Lya\ forest of \HeII\ down 
to $z \approx 2.55$ (G130M).  With the COS/G140L grating (one \HST\ orbit) we extended the \HeII\ measurements 
down to $z \approx 2.40$ at lower resolution (200~\kms).  We compared the \HeII\ absorbers with their 
\HI\ \Lya\ counterparts, using optical spectra taken by D. Reimers with the VLT/UVES (Ultraviolet Visual Echelle 
Spectrograph) in ESO program 071.A-0066 (see Shull \etal\ 2010).  We then used the $\eta$-ratio of \HeII/\HI\ 
absorption as a diagnostic of the metagalactic ionizing background spectrum.  

Figure 1 shows the observed COS (G130M and G140L) spectra of HE\,2347-4342, with strong absorption at 
wavelengths $\lambda \leq 1186$~\AA\ in the \Lya\ lines of \HeII\ (303.78~\AA\ rest frame).   
Figure~2 shows an expanded view of the COS/G130M spectra (1080--1200~\AA) relative to the extrapolated 
continuum.   Figure 3 shows the \HeII/\HI\ absorption ratio, calculated as the ratio of optical depths,
$\eta \equiv \tau_{\rm HeII} / \tau_{\rm HI}$.  Surprisingly, this AGN lacks a \HeII\ proximity effect.  Near the 
\HeII\ edge, between 1181 and 1186~\AA, we see strong \HI\ and \HeII\ absorption with inferred column 
densities $\log N_{\rm HI} \approx 16.29 \pm 0.05$ (Fox \etal\ 2008) and $\log N_{\rm HeII} \geq 16.3$ 
(Fechner \etal\ 2004).  There are also strong metal-line absorbers (\CIII, \CIV, \NV, \OVI) which can be 
modeled with photoionization codes.   These associated absorbers at $z = 2.885-2.905$ may have 
sufficient column densities to block some of the \HI\ and \HeII\ ionizing flux.  The estimated \HI\ column
density, $N_{\rm HI} \approx 2\times10^{16}$ cm$^{-2}$, corresponds to optical depth 
$\tau_{\rm HI} \approx 0.16$ at the (1 ryd) Lyman limit.  The \HeII\ column density could be much larger
($\eta_{\rm AGN} \approx 100$) and produce optical depth $\tau_{\rm HeII} \approx 4$ at 228~\AA\ (4~ryd).
Thus, it is possible that much of the \HeII-ionizing radiation is blocked near the AGN.  
As discussed in previous papers, these observations raise several puzzles about the structure of the AGN
environment.   Why does this powerful QSO produce no ``near-zone" ionization cavity?   How do the infalling 
associated absorbers ($z_a  > z_{\rm sys}$) overcome QSO radiation pressure?  If this QSO turned on in the
last Myr, its ionization front may still be propagating out of the dense circumnuclear gas that fueled the QSO 
in its early luminous phase.  

Many AGN are obscured, owing to inclination effects of the central accretion disk and dense gas in the 
galactic nucleus.  For example,  Seyfert galaxies were first classified as Type I or II, based on emission lines 
shown by their spectra.  However, these differences may arise from obscuring dust farther out from the 
accretion disc.   The composite UV/EUV spectra discussed in this paper were all Type I.   The absence 
of any photoelectric absorption edge at 912~\AA\ is consistent with clear escape channels for ionizing
radiation in the \HI\ Lyman continuum.    Still poorly constrained is the escape fraction of photons in the 
continuum of \HeII\ at 4-ryd (228~\AA).  While many \HeII\ quasars show proximity zones of various sizes
(Schmidt \etal\ 2018),  the large line-of-sight proximity zone of Q0302-002 (Syphers \& Shull 2014) is 
unusual compared to those in the two brightest \HeII\ quasars.  HS~1700+6416 has a small zone 
(Syphers \& Shull 2013), and HE~2347-4342 has no proximity zone at all, and shows evidence for 
associated absorption within the AGN (Figure 2).

\section{Analytic Photoionization Models}

We extend the formalism (Fardal \etal\ 1998) developed to assess the softening of the AGN ionizing spectrum 
as a result of radiative transfer through the IGM.  To quantify the effects of absorption internal to the AGN, we 
constructed a radiative equilibrium model with reprocessing by both IGM and internal AGN absorption.  
The emissivity of ionizing radiation ($j_{\nu}$) is produced by AGN sources with spectral index $\alpha_s$ and 
flux escape fractions, $f_{\rm esc}^{\rm (HI)}$ and $f_{\rm esc}^{(\rm HeII)}$ in the \HI\ and \HeII\ ionizing continua 
at 1 ryd and 4 ryd respectively.  We assume that \HI\ and \HeII\ are photoionized by metagalactic radiation with
a modified spectral index $\alpha_b$ characterizing the re-processed spectrum between 1--4 ryd (\HI\ and \HeII\
ionization edges).  The specific intensity $J_{\nu}$  is assumed to be in radiative equilibrium, so that the local 
source function, $S_{\nu} \equiv j_{\nu} / \kappa_{\nu} = J_{\nu}$.   The IGM opacity $\kappa_{\nu}$ comes from \HI\ 
absorbers with a distribution in \HI\ column density, $f(N) \propto N^{-\beta}$.  Appendix~A gives a full discussion 
of the integrated opacities of  \HI\ and \HeII.   In radiative equilibrium, the specific intensity 
$J_{\nu} = j_{\nu} / \kappa_{\nu}$ develops a softer spectrum with spectral index $\alpha_b > \alpha_s$.  
Using updated rates for \HI\ and \HeII\ photoionization and recombination, we derive a formula for $\alpha_b$ in 
terms  of $\alpha_s$, $\beta$, and the ratio $R \equiv f_{\rm esc}^{\rm (HI)} /  f_{\rm esc}^{(\rm HeII)}$. 

For the photoionization equilibrium in the low-density IGM, we assume Case-A radiative recombination rates to 
\HI\ and \HeII.  From these parameters, updated from those in Shull \etal\ (2010), we estimate the ratio of \HeII\  
and \HI\ column densities,
\begin{equation}
   \eta \equiv \frac {\rm N(He~II)}  {\rm N(H~I)} = \frac { n_{\rm HeIII} } {n_{\rm HII} } 
         \frac { \alpha_{\rm HeII}^{(A)} } {\alpha_{\rm HI}^{(A)} }
         \frac { \Gamma_{\rm HI} }  {\Gamma_{\rm HeII} }  
      \approx (1.77) \left[ \frac {J_{\rm HI}}  {J_{\rm HeII}} \right]
        \left[ \frac {3+\alpha_4} {3+\alpha_1} \right] T_{4.3}^{0.042} 
        \equiv C  \left[ \frac {J_{\rm HI}}  {J_{\rm HeII}} \right] \; .
\end{equation}
In this formula, we assume that H$^{+}$ and He$^{+2}$ are the dominant ionization stages of H and He, with 
$n_{\rm HeIII}/n_{\rm HII} \approx y  \equiv (Y/4)/(1-Y)$.  For the He/H ratio, we adopt $Y = 0.2477 \pm 0.0029$ 
by mass and $y = 0.0823$ by number (Peimbert \etal\ 2007) consistent with estimates of the primordial 
value, $Y_p = 0.2449 \pm 0.0040$ (Aver \etal\ 2015; Cyburt \etal\ 2016) and values from the CMB determined 
baryon density ({\it Planck} Collaboration 2015) and Big Bang nucleosynthesis (Cooke \& Fumagalli 2018).  
The accuracy of the He/H ratio demonstrates that it is not the limiting factor in modeling the spectral index 
from \HeII/\HI\ ratios. 

In equation (1), the quantities $\alpha_{\rm HI}^{(A)}$, $\alpha_{\rm HeII}^{(A)}$, $\Gamma_{\rm HI}$, and 
$\Gamma_{\rm HeII}$ are case-A radiative recombination rate coefficients and photoionization rates for \HI\ 
and \HeII.    We adopt the fits to radiative recombination rate coefficients from Shull \etal\ (2010) of  
$\alpha_{\rm HI}^{(A)} = (2.51 \times10^{-13}~{\rm cm}^3~{\rm s}^{-1}) T_{4.3}^{-0.736}$ and 
$\alpha_{\rm HeII}^{(A)} = (1.36 \times10^{-12}~{\rm cm}^3~{\rm s}^{-1}) T_{4.3}^{-0.694}$,
scaled to electron temperature $T_e = (10^{4.3}~{\rm K}) T_{4.3}$.   In the approximation that
photoionization cross sections for \HI\ and \HeII\ scale as $\nu^{-3}$ above threshold, we write
$\Gamma_{\rm HI} \approx 4 \pi J_{\rm HI} \sigma_{\rm HI} / h(3+\alpha_1)$ and 
$\Gamma_{\rm HeII} \approx 4 \pi J_{\rm HeII} \sigma_{\rm HeII} / h(3+\alpha_4)$.  
Here, $\sigma_{\rm HeII} / \sigma_{\rm HI} = 1/4$, and $J_{\rm HI}$ and $J_{\rm HeII}$ represent specific 
intensities of the ionizing radiation field at 1 ryd and 4 ryd.   The parameters $\alpha_1$ and $\alpha_4$ 
represent the spectral indices of the specific intensities at the ionization edges of  \HI\ (1 ryd) and \HeII\ 
(4~ryd) and they arise from the photoionization integrals for $\Gamma_{\rm HI}$ and $\Gamma_{\rm HeII}$.  
Hereafter, we adopt $C = 1.77$ with the assumptions that $\alpha_1 = \alpha_4$ and $T_{4.3} = 1$. 
The ratio is insensitive to $T_e$.  
 
The metagalactic ionizing background radiation is strongly influenced by intergalactic absorption 
(\Lya-forest absorbers) in the ionizing continua of both \HI\ ($h \nu \geq 13.6$~eV) and \HeII\  
($h \nu \geq 54.4$~eV).   There may also be internal opacity within the AGN sources, which we 
parameterize as flux escape fractions ($f_{\rm esc}$).   These internal opacities may be responsible
for the small \HeII\ proximity zones seen in some QSOs (Schmidt \etal\ 2018).  
To evaluate the radiative-transfer effects of QSO emission and IGM absorption, we assume that
IGM filtering results in a radiative source function $S_{\nu} = j_{\nu}/\kappa_{\nu}$, with emissivity 
$j_{\nu}$ and opacity $\kappa_{\nu}$.   The individual QSO sources are characterized by ionizing flux 
distributions, $F_{\nu} = F_0 (\nu/\nu_0)^{-\alpha_s}$, multiplied by escape fractions,
$f_{\rm esc}^{\rm HI}$ and $f_{\rm esc}^{\rm HeII}$ at 1~ryd and 4~ryd.   The distribution of \HI\ 
absorbers is expressed as $f(N_{\rm HI}) \propto N_{\rm HI}^{-\beta}$.   Studies of the \Lya\ forest with 
the VLT/UVES spectrograph (Kim \etal\ 2002) toward eight QSOs covering the \Lya\ forest at 
$1.5 < z  < 4$ found $\beta \approx 1.5$ over the range $12.5 < \log N_{\rm HI} < 15$.  A more recent 
survey (Rudie \etal\ 2013) of 15 AGN at $\langle z \rangle \approx 2.4$ with the {\it Keck}/HIRES 
spectrometer fitted the \Lya\ forest absorbers with a power-law index, $\beta = 1.65 \pm 0.02$ between 
$13.5 < \log N_{\rm HI} < 17.2$.   They also considered broken power-law distributions to better fit the 
incidence of strong absorbers with $\log N_{\rm HI} > 17.2$, the so-called Lyman-Limit systems  (LLS).  
The high-$N_{\rm HI}$ absorbers ($\log N_{\rm HI} \approx 15-17$) were fitted with a flatter index, 
$\beta \approx 1.48$.   We regard the slope ($\beta$) as a less constrained parameter, depending on 
which range of column densities is most relevant.  

In Appendix A, we describe analytic approximations to the integrated IGM opacities in the \Lya\ 
forest, using a single index ($\beta$) over appropriate ranges of $N_{\rm HI}$ or $N_{\rm HeII}$.     
This formulation neglects the opacity from the CGM of galaxies along the sightline.  These LLS are rare,
and their opacity is non-local, since they typically arise well outside the mean-free path of \HI-ionizing
photons, $\lambda_{\rm mfp} = 147 \pm 15$ physical Mpc (Rudie \etal\ 2013).  In addition, these strong 
\HI\ absorbers do not contribute appreciably to $\kappa_{\rm HeII}$, since the \HeII\ opacity occurs in 
absorbers with $\log N_{\rm HI} \approx 15-16$.
To a good approximation, we can analyze the ionization conditions in the diffuse IGM from the ratio 
of emissivities and opacities at the photoionizing continuum edges, 1 ryd (\HI) and 4 ryd  (\HeII),
\begin{equation}
      \frac {j_{\rm HeII}} {j_{\rm HI}} = \left[  \frac { f_{\rm esc}^{\rm (HeII)} }  { f_{\rm esc}^{(\rm HI)} } \right]
             4^{-\alpha_s}   \; \; {\rm and}   \; \;
      \frac {\kappa_{\rm HeII} } {\kappa_{\rm HI} } =  \left( \frac {\eta}{4} \right)^{(\beta-1)}   \; .
\end{equation}
Because the ionization states of \HI\ and \HeII\  are largely controlled by the ionizing intensities, $J_{\nu}$,
at the two edges, we define $J_{\rm HeII} / J_{\rm HI} \equiv 4^{-\alpha_b}$, where $\alpha_b$ is the
IGM-softened spectral index between 1 and 4 ryd.   Assuming radiative equilibrium, $J_{\nu} = S_{\nu}$.
Using equation (2) to evaluate the source function $j_{\nu} / \kappa_{\nu}$ at the two edges, we can 
express equation (1) as:
\begin{equation}
   \eta = C \; 4^{\alpha_s}  \left[  \frac { f_{\rm esc}^{\rm (HI)} }  { f_{\rm esc}^{(\rm HeII)} } \right]
    \left( \frac {\eta}{4} \right)^{(\beta-1)} \equiv C \; 4^{\alpha_b}   \; ,
 \end{equation} 
where $C \approx 1.77$.  For convenience, we define the escape-fraction ratio, 
$R =  f_{\rm esc}^{\rm (HI)} /  f_{\rm esc}^{(\rm HeII)}$.
In most cases we expect that $R \geq 1$, since the internal opacity of \HeII\ generally exceeds that of \HI.   
After taking the logarithm of both sides and substituting for $(\eta/4) =  4^{\alpha_b } \, (C/4)$, we arrive at 
an analytic  expression for the effective spectral index,
 \begin{equation}
   \alpha_b =   \frac {\alpha_s}{(2-\beta)} - \frac { (\beta-1) } { (2-\beta) } \left[ 1 - \frac {\ln C} {\ln 4} \right] 
       + \frac {1} {(2 - \beta)} \left[ \frac {\ln R}{\ln 4} \right]   \; .
\end{equation}
The last term on the right-hand side,
\begin{equation}
    \Delta \alpha_b  ^{\rm (int)} = \frac {1} {(2 - \beta)} \left[ \frac {\ln R}{\ln 4} \right]  \;   ,
\end{equation}
represents the additional spectral softening produced by internal absorption in the AGN.  This term 
will be zero ($R = 1$) if  the \HI\ and \HeII\ escape fractions are 100\% (or equal).

\section{Summary and Implications}   

A number of papers in the past two decades have calculated the production rates and cosmological 
radiative transfer of the photons that ionize gas in the IGM and CGM.  These include a series of models 
(Haardt \& Madau 1996, 2001, 2012;  Madau \& Haardt 2009; Fardal \etal\ 1998; Shull \etal\ 1999; 
Faucher-Gigu\`ere \etal\ 2009, 2020).    These calculations all assumed 100\% escape fractions in the
\HI\ and \HeII\ continua of Type~1 AGN and steep ionizing spectra with various spectral indices, 
$\alpha_s = 1.57$ (Madau \& Haardt 2009; Haardt \& Madau 2012), 
$\alpha_s = 1.6$ (Faucher-Gigu\`ere \etal\ 2009), $\alpha_s = 1.7$ (Faucher-Gigu\`ere 2020), and
$\alpha_s = 1.8$ (Fardal \etal\ 1998; Shull \etal\ 1999) based on previous AGN composite spectra 
from \HST/FOS (Zheng \etal\ 1997; Telfer \etal\ 2002).   The recent composite spectrum of 159~AGN 
measured by \HST/COS has a harder index, $\alpha_s = 1.41 \pm 0.15$ between 1--2 ryd.  Observations 
of \HeII/\HI\ ratios in QSO absorbers require an ionizing background considerably softer than an 
extrapolation of the spectrum from 1 ryd out to 4 ryd.   The intervening IGM may provide insufficient 
softening to reproduce the large \HeII/\HI\ ratios.  

We propose that many quasars have internal absorption in their \HeII\ continuum, to explain the large
\HeII/\HI\ ratios in the IGM and the \HeII\ proximity zones around many AGN.  Although wavelengths 
$\lambda < 228$~\AA\ are inaccessible to most QSO observations,  there is indirect evidence for 4-ryd flux
blockage from the small proximity zones of a number of AGN (Schmidt \etal\ 2018).   Observations of the 
distribution of the \HeII/\HI\  ratios, $\eta \equiv 1.77 \times 4^{\alpha_b}$, provide constraints on the 
metagalactic spectral index, $\alpha_b$, and the escape fraction of 4-ryd continuum photons.   To 
illustrate this quantitatively, we consider mean slopes, $\beta = 1.50$ and $\beta = 1.65$, of the \HI\ 
absorber distribution found in two recent surveys.   Equation (4) with $C = 1.77$ reduces to
\begin{eqnarray}
       \alpha_b &=& 2.00 \alpha_s - 0.588 + 1.443 \ln R ~~ ({\rm for~} \beta = 1.50) \\
       \alpha_b &=& 2.86 \alpha_s - 1.092 + 1.061\ln R ~~ ({\rm for~} \beta = 1.65)   \; . 
\end{eqnarray}
The \HeII/\HI\ ratio follows from $\eta = 1.77 \times 4^{\alpha_b}$.  Figure 4 shows curves of constant
$\eta$ on a two-parameter ($\alpha_s$, $\beta$) grid.  Recent determinations of  $\beta$ are marked  
as horizontal bands in blue ($\beta \approx 1.5$ from Kim \etal\ 2002) and red ($\beta = 1.65 \pm 0.02$
from Rudie \etal\ 2013).  The green dot and error bars show possible values for $\alpha_s = 1.41\pm0.15$ 
and $\beta = 1.65\pm0.02$, and the dashed rectangle covers the approximate range of $\eta$ observations 
in both HE~2347-4342 and HS~1700+6416.  Although the steeper slope ($\beta = 1.65 \pm 0.02$) 
proposed by Rudie \etal\ (2013) provides reasonable agreement with observed values of $\eta$, that
slope applies to their low-column \HI\ absorbers ($\log N_{\rm HI} < 15$).   Their sample at high \HI\
column densities is small, with just 11 and 12 absorbers in the last two bins 
($\log N_{\rm HI} \approx 16.5$ and 17.0).   

Appendix~A shows that most of the IGM opacity comes from strong absorbers, in the range known as
partial Lyman-Limit systems:  $\log N_{\rm HI} \approx 16-17.2$ for \HI\ at 1 ryd and
 $\log N_{\rm HI} = 15-16$ for \HeII\ at 4 ryd (assuming $\eta \approx 100$).  Over this high-column 
 range, Rudie \etal\ (2013) found that $\beta \approx 1.48-1.52$ provided a better broken power-law fit,
 with $\beta =1.66 \pm 0.03$ for the lower column \Lya\ forest.  For these reasons, we believe
 the blue band ($\beta \approx 1.5$) may be a better characterization of the opacity of the IGM required 
 for our local source-function modeling.   We now discuss the case for internal absorption.  
We assume that all of the \HI-ionizing radiation escapes the AGN ($f_{\rm esc}^{\rm (HI)} \approx 1$)
and explore \HeII\ escape fractions in the range $0.6-0.9$ ($R = 1.25-1.67$).  
For $\beta = 1.5$ (or $\beta = 1.65$) internal \HeII\ absorption would soften the spectral index of the 
metagalactic background by an additional $\Delta \alpha_b^{(\rm int)} \approx 0.32\,(0.46)$ for 
$f_{\rm esc}^{\rm (HeII)} = 0.8$ ($R = 1.25$) or by $\Delta \alpha_b^{(\rm int)}  \approx 0.74\,(1.05)$ for 
$f_{\rm esc}^{\rm (HeII)} = 0.6$ ($R = 1.67$).
These increases in background spectral index would correspond to ranges of $\eta \approx 60-110$ 
(for $\beta = 1.50$) and $\eta = 190-430$ (for $\beta = 1.65$).  

Our analytic formalism also places a \underline{lower limit} on the internal absorption at the \HeII\ edge,
for consistency with maximum values of $\eta$ and the relation $\eta \equiv 1.77 \times 4^{\alpha_b}$.
Adopting an upper bound of $\eta  < 200$ on \HeII /\HI,  we find $\alpha_b < 3.41$.  The additional spectral 
softening, $\Delta \alpha_b^{\rm (int)}$, from internal AGN absorption provides a lower limit on 
$f_{\rm esc}^{(\rm HeII)}$.   As shown in Table 2 for $\alpha_s = 1.4$, the effective indices with $R = 1$ 
(no internal AGN absorption) are $\alpha_b = 2.21$ ($\beta = 1.5)$ and $\alpha_b = 2.91$ ($\beta = 1.65)$.
Thus, the limit $\alpha_b < 3.41$ ($\eta < 200$) requires $\Delta \alpha_b^{\rm (int)} < 1.2$ ($\beta = 1.5$)
and $\Delta \alpha_b^{\rm (int)} < 0.5$ ($\beta = 1.65$).  From these limits, we infer that 
$R \equiv  f_{\rm esc}^{\rm (HI)} /  f_{\rm esc}^{(\rm HeII)} < 2.30$ and $f_{\rm esc}^{(\rm HeII)} > 0.43$
($\beta = 1.5$).  For $\beta = 1.65$, these limits are $R < 1.27$ and $f_{\rm esc}^{(\rm HeII)} > 0.78$
These values are illustrative, but they suggest that typical  \HeII\ escape fractions from AGN could
range from 50\% to 90\%.  Because individual AGN exhibit variations in spectral index about the mean 
value $\alpha_s \approx 1.4$ (Stevans \etal\ 2014), a few AGN could have smaller escape fractions, 
$f_{\rm esc}^{(\rm HeII)} < 0.4$.  These sources would produce \Lya\ absorbers with $\eta > 300$ in the 
proximity zone of the background source or near quasars transverse to the sight line.  
   
In addition to its effects on $\eta$-ratios, the harder QSO composite index, $\alpha_s \approx 1.4$,
could alter the thermal history of the IGM through photoelectric heating.  Appendix~B analyzes this
heating, showing that \HeII\ photoelectric heating augments the \HI\ heating by a constant factor (2.77)
independent of $\eta$.  This factor arises because a lower intensity at the 4~ryd edge decreases the
\HeII\ photoionization rate but increases the \HeII/\HI\ abundance ratio. A naive model, neglecting 
cosmological adiabatic expansion and balancing \HI\ photoelectric heating with cooling from radiative 
recombination and free-free emission, would predict large equilibrium temperatures, 
$T_{\rm eq}^{(H)} \approx  [(h \nu_1 / 1.13 k (\alpha_1+2)] \approx 40,000$~K, that are relatively 
insensitive to AGN spectral indices, $\alpha_1 \approx 1.4-1.7$.  The much lower inferred  \Lya\ absorber 
temperatures, $T \approx 12,000$~K at $z \approx 2.5$ (Becker \etal\ 2011; Hiss \etal\ 2019), likely 
result from non-equilibrium heating and cooling, including cosmological expansion and Compton 
cooling off the cosmic microwave background (McQuinn \& Upton Sanderbeck 2016).  
   
\noindent
The existence of internal AGN  \HeII\ (4-ryd) AGN absorption has several implications:
 \begin{enumerate}
 
 \item The observed absorption ratios (\HeII/\HI\  $\approx 50-200$) and mean AGN spectral index
       $\alpha_s \approx 1.4$ suggest that many AGN have partial flux blockage in their emergent
       4-ryd continuum, with $f_{\rm esc}^{(\rm HeII)} \approx 0.5-0.9$.  Internal \HeII\ absorption
       at $\lambda \leq 228$~\AA\ reduces the AGN production rate of \HeII\ continuum photons 
       and shifts the \HeII\ epoch of reionization to lower redshifts.  
        
 \item An AGN absorption edge at 54.4~eV, together with flux recovery at higher energies, will
        affect photoionization models of heavy elements with photoionization thresholds between 
        3--10 ryd, including effects of cosmological radiative transfer.  These effects are significant for
        ions such as  \CIV\ (47.89~eV), \CV\ (64.49~eV),  \NIV\ (47.45~eV), \NV\ (71.47~eV),
         \OIII\ (54.94~eV), as well as higher ions \OVI\ and \NeVIII.  
        
 \item Internal AGN absorption at 4~ryd may be responsible for the small \HeII\ proximity regions
           seen in some AGN.  The variable sizes of these zones depends on variations in both AGN
           intrinsic spectral index ($\alpha_s$) and escape fraction ($f_{\rm esc}$).   Separating these 
           effects will require observing large numbers of QSOs to measure the distribution of $\alpha_s$ 
           and $\eta$ and searching for correlation of $f_{\rm esc}$ with the size of proximity zones. 
         
 \item Our analytic model for spectral softening by IGM and AGN absorption uses mean values of
           parameters ($\alpha_s$, $\beta$, $f_{\rm esc}$).  Numerical simulations could improve the 
           model by selecting AGN with a range of spectral indices and escape fractions.   These 
           parameters probably depend on properties of the AGN and its host galaxy and vary with redshift.
          
 \end{enumerate}

\acknowledgements

\noindent
We acknowledge helpful discussions with Piero Madau and Mark Giroux on \HeII\ absorption
and the possible effects of AGN internal absorption.  We also acknowledge the useful questions
posed by the referee about our analytic model and effects of the harder AGN spectral index on 
IGM thermal history.   This research was supported by grants HST-GO-13301 and 
HST-GO-15084 from the Space Telescope Science Institute to the University of Colorado Boulder. 

\newpage

\appendix


\section{Appendix A:  IGM Opacity of \HI\ and \HeII}

As shown below, the opacity at 1 ryd in the photoionized IGM is dominated by \Lya-forest 
absorbers with continuum optical depths $\tau_{\rm HI} \approx 0.1-1$ 
($\log N_{\rm HI} \approx 16.2-17.2$), whereas the \HeII\ opacity at 4~ryd comes from absorbers 
with $\log N_{\rm HI} \approx 15-16$.  This lower range arises from the large \HeII\ abundance 
(\HeII/\HI\ $\approx 50-200$).   At the photoionization continuum edge of \HI, the absorption cross 
section is $\sigma_{\rm HI} = 6.30 \times 10^{-18}$~cm$^{2}$.  The \HeII\ photoionization cross 
section at 4~ryd is four times smaller, $\sigma_{\rm HeII} = 1.58 \times 10^{-18}$~cm$^{2}$.  
The column densities, $N_i = \sigma_i^{-1}$, needed to produce continuum optical depths 
$\tau_i =1$ are $N_{\rm HI} = 1.59 \times 10^{17}$~cm$^{-2}$ and 
$N_{\rm HeII} = 6.35 \times 10^{17}$~cm$^{-2}$.  

The opacity ratio, $\kappa_{\rm HeII} / \kappa_{\rm HI}$, therefore samples the \Lya\ forest at 
different portions of the column-density distribution, expressed here as a power law, 
$f(N_{\rm HI}) = C_H N_{\rm HI}^{-\beta}$.   From their {\it Keck}/HIRES spectroscopic survey 
probing the IGM at $\langle z \rangle = 2.4$, Rudie \etal\ (2013) found that a single index 
$\beta = 1.65\pm0.02$ provides a reliable fit for $13.5 < \log N_{\rm HI} < 17.2$ (see their Figure 2).   
They noted that ``a single power law reproduces the distribution with reasonable fidelity" and that 
``no strong breaks in the distribution are evident."  They also considered broken power-laws to fit 
the incidence of high column density LLS absorbers (O'Meara \etal\ 2013), particularly those with 
$\log N_{\rm HI} > 18.0$.   By separately considering pathlengths through the CGM of galaxies located
within 300 physical kpc of the sight line, they found additional opacity with power-law indices 
$\beta \approx 1.42-1.45$.  Although these broken power-law fits provide more accurate total opacities, 
we only consider \Lya\ absorbers with $\log N_{\rm HI} \leq 17.2$ to describe radiative conditions in the 
diffuse IGM.  We do so for several reasons.  First, the source function, $S_{\nu} = j_{\nu} / \kappa_{\nu}$, 
must reflect local conditions, not opacities from rare, more distant LLS.  Second, the higher $N_{\rm HI}$ 
absorbers are associated with intervening galaxies and overdense portions of the CGM.    Third, 
strong absorbers with $\log N_{\rm HI} > 18$ are self-shielding in the Lyman continuum ($\tau \gg 1$) 
and do not contribute appreciably to the source function.  Thus, in the diffuse IGM at $z < 3$, the 
ionization conditions are dominated by quasars, whose ionizing radiation escapes their host galaxy.    

For Poisson distributed absorbers with \HI\ column densities $N_1 \leq N_{\rm HI} \leq N_2$, the mean
 \HI\ (1 ryd) opacity of the IGM can be expressed 
(Paresce \etal\ 1980) as 
\begin{equation} 
   \kappa_{\rm HI}   = \int_{\rm N_1}^{\rm N_2}  f(N_{\rm HI} )
         \left[  1 - \exp(-\tau_{\rm HI})  \right] \; dN_{\rm HI}  \; ,     
 \end{equation} 
 where $f(N_{\rm HI})$ is the distribution of \HI\ absorbers and $\tau_{\rm HI} = \sigma_{\rm HI} N_{\rm HI}$ 
 is the \HI\ continuum optical depth at 1 ryd.   We integrate the opacity from $\log N_1 = 13.0$ to 
 $\log N_2 = 17.2$ ($N_2 \gg N_1$).   Since $\tau_{\rm HI}  < 1$, we approximate the factor 
 $[1 - \exp(-\tau_{\rm HI})] \approx \tau_{\rm HI}$ to find
  \begin{equation}
       \kappa_{\rm HI} \approx \sigma_{\rm HI} C_{\rm HI}   \int_{\rm N_1}^{N_2} N_{\rm HI}^{(-\beta+1)} 
               \;  dN_{\rm HI}  \approx \frac {C_{\rm HI}  \, \sigma_{\rm HI}} {(2-\beta)}  
      \left[ N_2^{(2-\beta)} - N_1^{(2-\beta)} \right]  
             \approx \frac {C_{\rm HI}  \, \sigma_{\rm HI}} {(2-\beta)} N_2^{(2-\beta)}  \; . 
  \end{equation}
 For observed values, $\beta \approx 1.4-1.7$, the \HI\ opacity integral is dominated by absorbers 
 at the upper end of the distribution ($\tau \approx 0.1-1$).  For example, 64\% of the opacity (for 
 $\beta = 1.65$) comes from ``partial LLS" with $16.0 \leq N_{\rm HI} < 17.2$.   Extending the 
 integration to LLS with $17.2 < \log N_{\rm HI} < 18.0$ would provide an additional 39\% to the opacity.  
 However, these LLS are associated with galaxies and their CGM, and they would be opaque in
 both continua ($\tau_{\rm HI} \gg 1$ and $\tau_{\rm HeII} \gg 1$).  In our source-function approach 
 to the ionizing radiation  and \HeII/\HI\ observations, we concentrate on absorbers in the diffuse 
 IGM that are optically thin in 1 ryd and 4 ryd continua.  
  
To evaluate the mean \HeII\ opacity at 4 ryd in the optically thin \Lya\ forest, we adopt a constant 
abundance ratio, $\eta = N_{\rm HeII} / N_{\rm HI}$, and a  ratio of photoionization cross sections, 
$\sigma_{\rm HeII} = \sigma_{\rm HI}/4$, Because $\eta \gg 1$, the upper limit of the \HeII\ integral,
$\sigma_{\rm HeII}^{-1}$, corresponds to a smaller \HI\ column density,
$N^*_2 = 4N_2 /\eta = (10^{15.8}~{\rm cm}^{-2})(\eta/100)^{-1}$, at which $\tau_{\rm HeII} = 1$,
\begin{equation} 
 \kappa_{\rm HeII} =  \left[  \frac {\eta \, \sigma_{\rm HeII} \, C_{\rm HI} } {4} \right]
        \int_{\rm N_1}^{N^*_2}  N_{\rm HI}^{(-\beta+1)} \; dN_{\rm HI}
    \approx  \left[ \frac {\eta \, C_{\rm HI} \, \sigma_{\rm HI} } { 4 \, (2-\beta)} \right] (N^*_2)^{(2-\beta)}  \;  .
\end{equation} 
With these approximations,  we can express the ratio of opacities as
\begin{equation}
   \frac { \kappa_{\rm HeII} } {\kappa_{\rm HI} } =  \left( \frac {\eta}{4} \right) 
              \left[ \frac {N^*_2} {N_2} \right]^{(2-\beta)} 
       = \left( \frac {\eta} {4} \right)^{(\beta-1)}   \; \; .
\end{equation}
This analytic relation is employed in equation (2), and it assumes that the local opacities in the
IGM are governed by \Lya\ absorbers up to $\log N_{\rm HI} = 17.2$.  This formulation neglects 
the additional opacity from the CGM of galaxies along the sightline.  Because these LLS are rare,
their opacity is non-local and their numbers are stochastic.  In addition, these strong \HI\ absorbers 
do not contribute appreciably to $\kappa_{\rm HeII}$, since the upper limit of integration for 
\HeII\ ($\tau_{\rm HeII} \approx 1$) only extends to $\log N_{\rm HI} \approx 15.8$ for $\eta = 100$,
where a single index $\beta$ is a good fit to the distribution of column densities.

 
 \section{Appendix B: Thermal Equilibrium with H~I and He~II Photoionization}

In this Appendix, we compute the photoelectric heating of \HI\ and \HeII\ and estimate equilibrium 
temperatures, when balanced against radiative cooling from recombination and free-free emission 
from H$^+$ and He$^{+2}$.   We also consider other sources of cooling, including cosmological 
adiabatic expansion and collisional excitation of \Lya\ and metal forbidden-lines. The IGM absorbers 
are not in equilibrium, and adiabatic cooling can be effective at $z > 2$ (Hui \& Gnedin 1997). 
Compton cooling by photons in the cosmic microwave background can be a minor source
(McQuinn \& Upton Sanderbeck 2016).  

The photoionization rates of \HI\ and \HeII\ can be approximated as
\begin{eqnarray} 
   \Gamma_{\rm HI}    & \approx & \int_{\nu_1}^{\infty} \frac {J_{\nu}}{h \nu} \sigma_{H,0} 
                                                     \left( \frac {\nu}{\nu_1} \right)^{-3} \, d\nu 
                \equiv \frac { J_1 \sigma_1}   {h (\alpha_1 + 3)}  \\
                  \\
   \Gamma_{\rm HeII} & \approx & \int_{\nu_2}^{\infty} \frac {J_{\nu}}{h \nu} \sigma_{He,0} 
                                                    \left( \frac {\nu}{\nu_4} \right)^{-3} \, d\nu 
                 \equiv \frac { J_4  \sigma_4}  {h (\alpha_4 + 3)} \; \; .                      
\end{eqnarray}
We approximate the photoionization cross sections as $\sigma_{\rm i} (\nu/\nu_i)^{-3}$ 
and define the cross sections 
$\sigma_1 \equiv \sigma_{H,0} =  6.304 \times 10^{-18}$~cm$^2$ and
$\sigma_4 \equiv \sigma_{He^+,0} = 1.576 \times 10^{-18}$~cm$^2$ at ionization threshold 
energies $h\nu_1 = 13.598$~eV (\HI) and $h\nu_4 = 54.418$~eV (\HeII).  
Here, $J_{\nu} = 4 \pi I_{\nu}$ is the specific intensity (erg~cm$^{-2}$~s$^{-1}$~Hz$^{-1}$)
integrated over $4\pi$ sterradians, and $\alpha_1$ and $\alpha_4$ are the spectral indices of 
the continuum at 1~ryd and 4~ryd, respectively.   After IGM filtering and reprocessing, these 
indices are uncertain, but we assume they are the same.  For highly ionized gas with 
$n_{\rm He} / n_{\rm H} = 0.0823$, the electron density $n_e \approx 1.165 n_H$.  
If the \Lya\ absorbers are in photoionization equilibrium with radiative recombination,
$n_{\rm HI} \Gamma_{\rm HI} = n_e n_{\rm HI} \alpha_H^{(1)}$, and
\begin{equation}
        n_{\rm HI} = 1.165 n_H^2 \alpha_H^{(1)} / \Gamma_{\rm HI} \approx 
                 \frac {1.165 \, h \, n_H^2 \alpha_H^{(1)} (\alpha_1 + 3) }  { J_1 \sigma_1 }  \;   .
\end{equation} 
We use the case-A recombination rate coefficient, $\alpha_H^{(1)}$, for low-density absorbers, 
which are optically thin to Lyman continuum photons from recombinations to the ground state.  

The ionizing radiation injects photoelectrons into the gas, heating the gas at rates
$n_{\rm HI} {\cal H}_{\rm HI}$ and $n_{\rm HeII} {\cal H}_{\rm HeII}$ per unit volume,
with rate coefficients (erg~cm$^3$~s$^{-1}$)
\begin{eqnarray} 
   \cal{H}_{\rm HI}    & \approx &  \int_{\nu_1}^{\infty} \frac {J_{\nu} } { h \nu } \sigma_{\nu } 
                                                     \left( \frac {\nu}{\nu_1} \right)^{-3} (h\nu - h\nu_1) \, d\nu 
                = \frac {J_1 \, \sigma_1 \,  \nu_1} {(\alpha_1 + 2) (\alpha_1 + 3)}  \\                  
   \cal{H}_{\rm HeII} & \approx &  \int_{\nu_4}^{\infty} \frac {J_{\nu}} {h \nu } \sigma_{\nu } 
                                                    \left( \frac {\nu} {\nu_0} \right)^{-3} (h\nu - h\nu_4) \ \, d\nu 
                 = \frac {J_4 \, \sigma_4 \, \nu_4} {(\alpha_4 + 2)(\alpha_4 + 3)} \; \; .                      
\end{eqnarray}
The mean photoelectron energy per ionization, $({\cal H}_i / \Gamma_i) = h \nu_i / (\alpha_i+2)$, 
reflects the fact that photoionization cross sections drop off rapidly above the ionization edges.  
From the above equations, we find a constant ratio of \HeII\ to \HI\ photoelectric heating rates,
\begin{equation}
   \frac {n_{\rm HeII} \, {\cal H}_{\rm HeII}} {n_{\rm HI} \, {\cal H}_{\rm HI}} = \eta \left( \frac {J_4}{J_1} \right)
             \left[ \frac {(\alpha_1 + 2)(\alpha_1 + 3)} {(\alpha_4 + 2)(\alpha_4 +3)} \right] \approx 1.77 \; ,
\end{equation}
the result of the three relations:
$\sigma_4 \, \nu_4 = \sigma_1 \, \nu_1$, $(J_4/J_1) = 4^{-\alpha_b}$, $\eta = (1.77)4^{\alpha_b}$ 
and the assumption that the term inside brackets is unity.   
The \HeII\ photoelectric heating augments the \HI\ heating by a constant factor (2.77); 
a lower intensity at the 4~ryd edge decreases $\Gamma_{\rm HeII}$ but increases the \HeII/\HI\ 
abundance ratio ($\eta$).  For the same reasons, the relative values of \HeII\ and \HI\ heating 
should not change with moderate internal AGN absorption.
            
For a pure hydrogen plasma in thermal equilibrium, ignoring adiabatic expansion or Compton 
cooling, we balance photoelectric heating (eq.\ [B5]) with cooling from radiative recombination and 
free-free emission, expressed as $n_e n_{\rm HII} \alpha_H^{(1)} (1.13 kT)$.  The factor 1.13 
roughly accounts for free-free cooling at $T \approx 10^4$~K.   Substituting for $n_{\rm HI}$ 
(eq.\ [B4]), we find a large equilibrium temperature, 
\begin{equation}
     T_{\rm eq}^{(H)} \approx  \left[ \frac { (h \nu_1 /k) } {1.13  (\alpha_1+2) } \right] 
            \approx 41,000~{\rm K}    \;  ,
\end{equation}    
for the QSO spectral index $\alpha_1 = 1.4$ from the \HST/COS composite spectrum 
(Stevans \etal\ 2014).   The temperature drops slightly for the softer spectral indices:  
39,500~K for $\alpha_1 = 1.57$ (Telfer \etal\  2002) and 
37,000~K for $\alpha_1 = 1.70$ (Lusso \etal\ 2015).  Including \HeII\ photoelectric boosts 
the heating from \HI\ alone by a factor of 2.77, and \HeIII\ radiative raises the cooling by a
factor of 1.45.  This would lead to an even higher characteristic temperature,
\begin{equation}
     T_{\rm eq} ^{(H,He)} \approx  \left[ \frac {(2.77/1.45) (h \nu_1 /k) } {1.13  (\alpha_1+2) } \right] 
            \approx 78,700~{\rm K}    \; .    
\end{equation}    
However, significant cooling comes from cosmological adiabatic expansion, at a rate per unit 
volume, ${\cal L}_{\rm ad} = 3 H(z) nkT$.  The Hubble expansion parameter at $z > 2$ can be 
approximated as  $H(z) \approx H_0 \Omega_m^{1/2} (1+z)^{3/2}$.  We adopt a total particle 
density $n = 2.247 n_H$, with mean IGM density 
$\overline{n}_H(z) = (1.88 \times 10^{-7}~{\rm cm}^{-3})(1+z)^3$ and baryon 
overdensity factor $\Delta_b$.  The ratio of adiabatic cooling to photoelectric heating is then, 
\begin{equation}
       \frac { (3nkT) H(z) } { 2.77 \, n_{\rm HI} \,  {\cal H}_{\rm HI} } \approx 
              (1.12) \left[ \frac {1+z} {3.5} \right]^{-3/2}  \Delta_b^{-1} \, T_4^{1.736}  \; ,
\end{equation}
a significant contribution at  $z \approx 2.5$, although less so for \Lya\ absorbers with 
$\Delta_b > 10$.  Overdense absorbers will experience relatively more photoelectric heating, 
since $n_{\rm HI} \propto n_H^2$ whereas adiabatic cooling is linear with $n_H$.  

Most \Lya\ absorbers have insufficient temperatures to produce significant collisional excitation 
of the first excited state of hydrogen at 10.2~eV.  At $z \approx 2-3$, temperatures inferred 
from \Lya\ line widths and curvature (Becker \etal\ 2011; Hiss \etal\ 2019) suggest that
$T \approx 12,000$~K, too low for significant \Lya\ cooling.  Collisional excitation of forbidden 
lines of trace metal ions will contribute some cooling.  However, at typical  \Lya-forest 
metallicities ($0.01Z_{\odot}$), cooling from [\OII] and [\OIII] at 12,000~K is only 10\% that from 
hydrogen radiative recombination and free-free emission.


\small{

}



\begin{deluxetable}{lrrrrr}
\tablecolumns{6}
\tabletypesize{\footnotesize}
\tablenum{1}
\tablewidth{0pt}
\tablecaption{\HST/COS Observation Details\tablenotemark{a} }

\tablehead{
\colhead{Grating} 
&\colhead{Setting} 
&\colhead{Range} 
&\colhead{Exp.\  (s)}
&\colhead{Obs.\ Date}
&\colhead{Obs.\ No.} 
 }
         
\startdata 
    G130M  &  1222~\AA\  & 1067 - 1363~\AA\   & 14813.76  & 2014 July 08   &  lc8i01010  \\
    G130M  &  1222~\AA\  & 1067 - 1363~\AA\   & 14813.76  & 2014 July 12   &  lc8i03010  \\
    G130M  &  1222~\AA\  & 1067 - 1363~\AA\   & 11766.66  & 2014 Aug 28   &  lc8i02010  \\
    G140L   &  1105~\AA   & 1118 -  2251~\AA\  &   2920.29  & 2014 Aug 28   &  lc8i02020  
\enddata 

\tablenotetext{a} {Observations of the bright \HeII\ quasar HE~2347-4342 were taken under 
HST GO Program ID 13301 (PI Shull) using COS gratings, as listed.  }

\end{deluxetable}



\begin{deluxetable}{lcccc}
\tablecolumns{5}
\tabletypesize{\footnotesize}
\tablenum{2}
\tablewidth{0pt}
\tablecaption{Softened Spectral Index and \HeII/\HI\ Ratio }

\tablehead{
\colhead{AGN Index}
&\colhead{ ($\alpha_b$, $\eta$)\tablenotemark{a} }
&\colhead{ ($\alpha_b$, $\eta$)\tablenotemark{a} }
&\colhead{ ($\alpha_b$, $\eta$)\tablenotemark{a} }
&\colhead{ ($\alpha_b$, $\eta$)\tablenotemark{a} } \\

\colhead{ }
& \colhead{$\beta=1.40$ }
& \colhead{$\beta=1.45$ } 
& \colhead{$\beta=1.50$ } 
& \colhead{$\beta=1.65$ } 
 }

             
\startdata
 $\alpha_s = 1.3$ &  (1.78, 21)  & (1.88, 24)  &  (2.01, 29)   &  (2.67, ~67)  \\
 $\alpha_s = 1.4$ &  (1.94, 26)  & (2.06, 31)  &  (2.21, 38)   &  (2.91, 100)  \\
 $\alpha_s = 1.5$ &  (2.11, 33)  & (2.25, 40)  &  (2.41, 50)   &  (3.19, 148)  \\
 $\alpha_s = 1.6$ &  (2.28, 42)  & (2.43, 51)  &  (2.61, 66)   &  (3.48, 220)  \\
 $\alpha_s = 1.7$ &  (2.44, 52)  & (2.61, 66)  &  (2.81, 81)   &  (3.77, 327)  
\enddata 

\tablenotetext{a} {Values of softened spectral index  ($\alpha_b$) and corresponding \HeII/\HI\ ratio,
$\eta = 1.77 \times 4^{\alpha_b}$ determined from eq.~(4) for several values of AGN intrinsic spectral 
index ($\alpha_s = 1.3-1.7$) and \Lya-forest column density distributions ($\beta = 1.40$, 1.45, 1.50, 1.65).  
These values neglect the escape fraction term by setting $R = 1$ in eq.\ (4).   Internal He~II absorption within 
the AGN would further soften the background spectral index, by $\Delta \alpha_b = [\ln R / \ln 4] / (2 - \beta)$,
where $R =  f_{\rm esc}^{\rm (HI)} / f_{\rm esc}^{\rm (HeII)}$.  }

\end{deluxetable}



\begin{figure}[h]
\includegraphics[angle=0,scale=0.67]{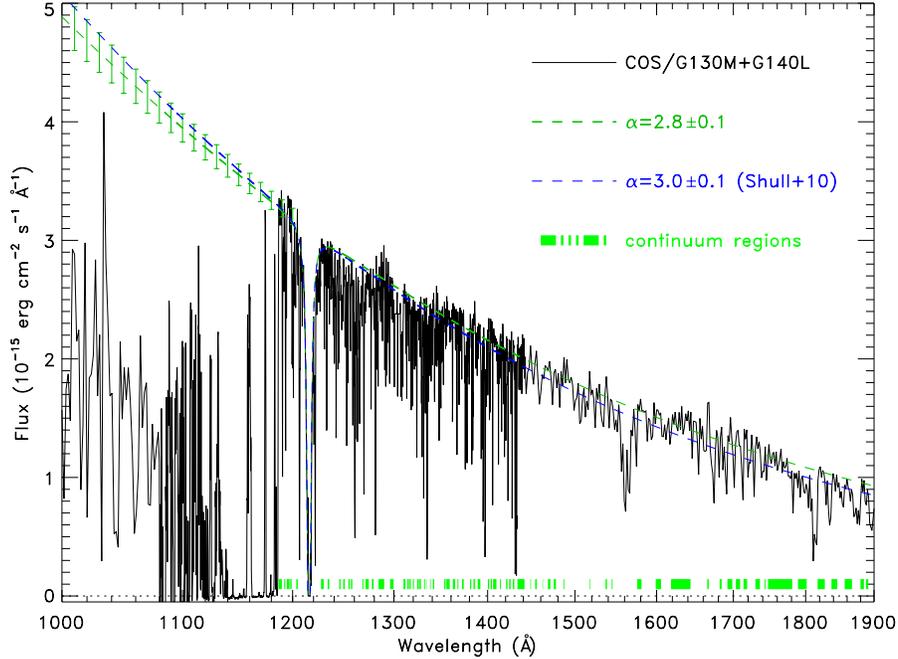}
\caption{\small{Far-UV spectrum of the quasar HE\,2347-4342 at redshift $z_{\rm sys} = 2.886 \pm 0.001$
taken by the Cosmic Origins Spectrograph aboard \HST.  The central portion (1080--1200~\AA) was obtained
using the G130M grating in the 1222~\AA\ (super-blue) setting.   Spectra from 1400--1900~\AA\ and at
$\lambda < 1100$~\AA\ were taken with the low-resolution G140L grating.  The G140L data were used 
to define the continuum, $F_{\lambda} \propto \lambda ^{-\alpha}$ with $\alpha = 2.8\pm0.1$, slightly different 
than the previous fit ($\alpha = 3.0 \pm 0.1$) in Shull \etal\ (2010).  Strong Gunn-Peterson absorption from
\HeII\ \Lya\ lines appears at wavelengths $\lambda < 1190$~\AA, recovering gradually at shorter wavelengths. }
}
\end{figure}



\begin{figure}[h]
\includegraphics[angle=0,scale=0.67]{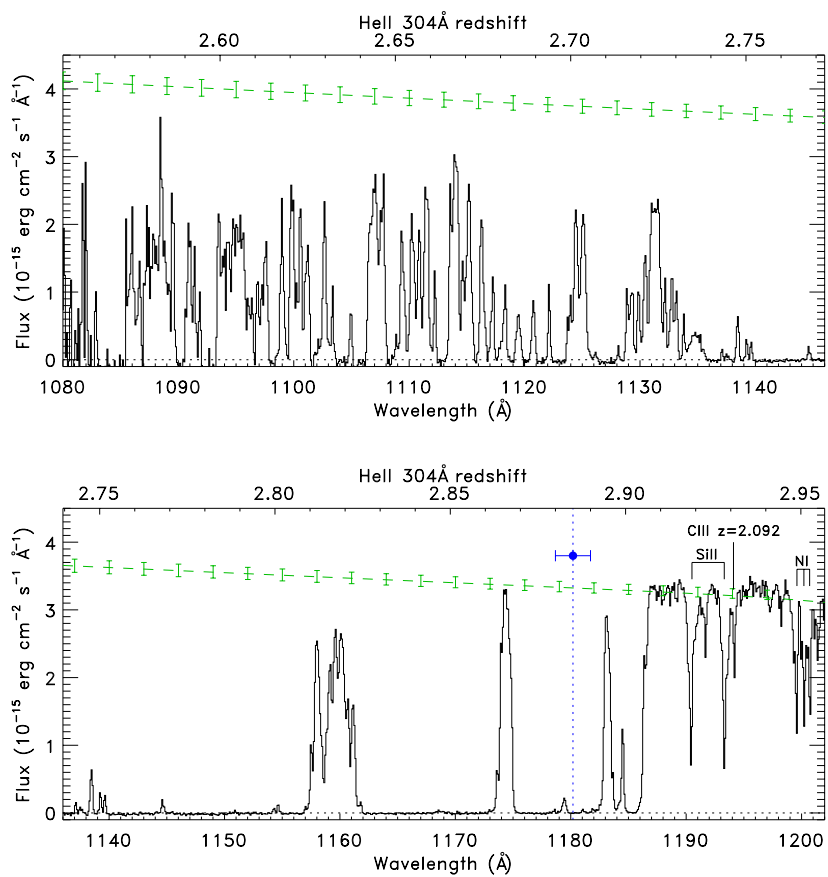}
\caption{\small{ COS/G130M spectrum of HE\,2347-4342 showing  \HeII\ absorption troughs (1140-1186~\AA) and 
flux transmission recovery from 1140~\AA\  down to 1080~\AA\ ($z = 2.56-2.75$).  The systematic redshift of the AGN
($z_{\rm sys} = 2.886\pm0.001$) is marked with vertical  dashed line.  Absorption from 1181-1190~\AA\ comes from 
associated absorbers at velocities up to 1480~\kms\ redward of $z = 2.886$. }
}
\end{figure}



\begin{figure}[h]
\includegraphics[angle=0,scale=0.70]{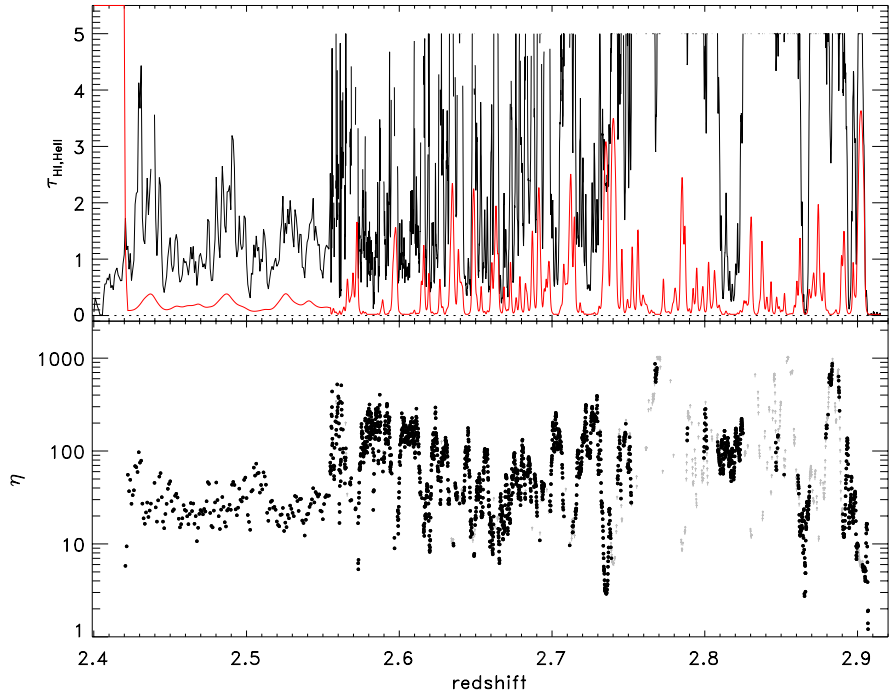}
\caption{\small{Comparison of \Lya\ line absorption for \HeII\ and \HI\ toward HE~2347-4342 at redshifts
$z = 2.4-2.9$.  Top panel shows the \Lya\ optical depths for \HeII\ (black, stronger values) and \HI\ (red, weaker).
Bottom panel shows the $\eta$-ratio of \HeII/\HI\ optical-depths (eq.\ [1]) which depends sensitively on the 
spectrum of the IGM-filtered metagalactic background at 1-5 ryd .  }
}
\end{figure}



\begin{figure}[ht]
\includegraphics[angle=0,scale=1.0]{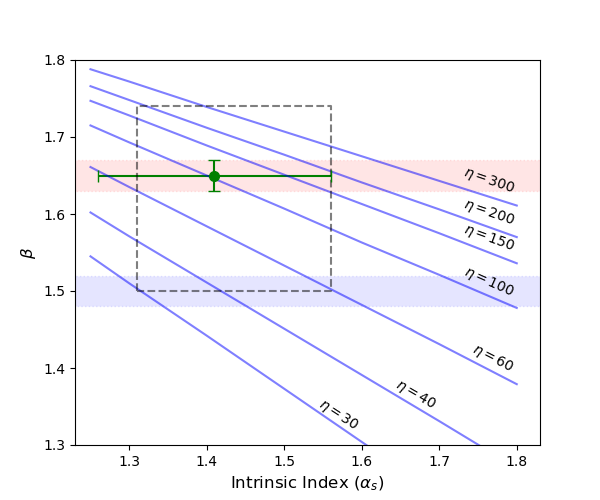}
\caption{\small{ Values of constant $\eta = N_{\rm HeII} / N_{\rm HI}$ ranging from
30 to 300 on a grid of AGN intrinsic spectral index ($\alpha_s$) and slope ($\beta$) 
of \HI\ column-density distribution.   Recent determinations of  $\beta$ are shown 
as horizontal colored bands in red ($\beta = 1.65\pm0.02$ from Rudie \etal\ 2013) 
and blue ($\beta \approx 1.5$ from Kim \etal\ 2002). The dashed rectangle covers the 
approximate range of $\eta$ observations in both HE~2347-4342 and HS~1700+6416.
The green dot and error bars show best-fit values for $\alpha_s = 1.41\pm0.15$ 
(Stevans \etal\ 2014) and $\beta = 1.65\pm0.02$ (Rudie \etal\ 2013).  However,
$\beta = 1.5$ may be more appropriate for the strong absorbers that dominate the
opacity. }
}
\end{figure}



\begin{references}

\reference{}  Aver, E., Olive, K. A., \& Skillman, E. D. 2015, JCAP, 7, 011 
\reference{}  Becker, G. D., Bolton, J. S., Haehnelt, M. G., \& Sargent, W. L. W. 2011,
     \mnras, 410, 1096
\reference{}  Bian, F., \& Fan, X. 2020, \mnras, 493, L69
\reference{}  Boksenberg, A. \& Sargent, W. L. W. 2015, \apjs, 218, 93  
\reference{}  Cooke, R. J., \& Fumagalli, M. 2018, Nature Astronomy, 2, 957 
\reference{}  Cyburt, R. H., Fields, B. D., Olive, K. A., \& Yeh, T.-H. 2016, RMP, 88, 015004 
\reference{}  Dall'Aglio, A., Wisotzki, L., \& Worseck, G. 2008a, A\&A, 480, 459 
\reference{}  Dall'Aglio, A., Wisotzki, L., \& Worseck, G. 2008b, A\&A, 491, 465 
\reference{}  Fardal, M. A., Giroux, M. L., \& Shull, J. M.  1998, \aj, 115, 2206 
\reference{}  Faucher-Gigu\`ere, C.-A., Lidz, A., Zaldarriaga, M., \& Hernquist, L. 2009, \apj, 703, 1416 
\reference{}  Faucher-Gigu\`ere, C.-A. 2020, \mnras, 493, 1614 
\reference{}  Fechner, C., Baade, R., \& Reimers, D.  2004, A\&A, 418, 857 
\reference{}  Fechner, C., Reimers, D.,  Kriss, G. A., \etal\ 2006, A\&A, 455, 91
\reference{}  Finkelstein, S. L., D'Aloisio, A., Paardekooper, J.-P., \etal\ 2019, \apj, 879, 36 
\reference{}  Fox, A., Bergeron, J., \& Petitjean, P.  2008, \mnras, 388, 1557 
\reference{}  Graziani, L., Maselli, A., \& Maio, U. 2019, \mnras, 482, L112 
\reference{}  Green, J. C., Froning, C. S., Osterman, S., \etal\ 2012, \apj, 744, 60 
\reference{}  Gunn, J. E., \& Peterson, B. A. 1965, \apj, 142, 1633 
\reference{}  Haardt, F., \&  Madau, P. 1996, \apj, 461, 20 
\reference{}  Haardt, F., \& Madau, P. 2001, in Clusters of Galaxies and the High Redshift 
    Universe Observed in X-rays, ed. D. M. Neumann \& J. T. V. Tran (Saclay: CEA), 64 
\reference{}  Haardt, F., \& Madau, P. 2012, \apj, 746. 125 
\reference{} Hiss, H., Walther, M., O\~norbe, J., \& Hennawi, J. F. 2019, \apj, 876, 71
\reference{}  Hui, L., \& Gnedin, N. Y. 1997, \mnras, 292, 27  
\reference{}  Izotov, Y. I., Schaerer, D., Thuan, T. X., \etal\ 2018, \mnras, 461, 3683 
\reference{}  Jakobsen, P., Jansen, R. A., Wagner, S., \& Reimers, D. 2003, A\&A, 397, 891 
\reference{}  Kim, T.-S.,  Carswell, R. F., Cristiani, S., D'Odorico, S., \& Giallongo, E. 
    2002, \mnras, 335, 555 
\reference{}  Kriss, J., Shull, J. M., Oegerle, W., \etal\, 2001, Science, 293, 1112 
\reference{}  Lusso, E., Worseck, G., Hennawi, J. F., \etal\ 2015, \mnras, 449, 4204 
\reference{}  Madau, P., \& Dickinson, M. 2014, \araa, 52, 415 
\reference{}  Madau, P., \& Haardt, F. 2009, \apj, 693, L100 
\reference{}  Madau, P., \& Haardt, F. 2015, \apj, 813, L8 
\reference{} McQuinn, M., \& Upton Sanderbeck, P. R. 2016, \mnras, 456, 47 
\reference{}  Miralda-Escud\'e, J., Cen, R.,  Rauch, M., \& Ostriker, J. P. 1996, \apj, 471, 582  
\reference{}  O'Meara, J. M., Prochaska, J. X., Worseck, G., Chen, H.-W., \& Madau, P.
                 2013, \apj, 765, 135
\reference{}  Paresce, F., McKee, C. F., \& Bowyer, S. 1980, \apj, 240, 387 
\reference{}  Peimbert, M., Luridiana, V., \& Peimbert, A. 2007, \apj, 666, 636 
\reference{}  Planck Collaboration \etal\ 2015, A\&A, 594, A13 
\reference{}  Puchwein, E., Bolton, J. S.,  Haehnelt, M. G., \etal\ 2015, \mnras, 450, 4081 
\reference{}  Puchwein, E., Haardt, F., Haehnelt, M. G., \& Madau, P. 2019, \mnras, 485, 47 
\reference{}  Reimers, D.,  K\"ohler, S., Wisotzki, L., \etal\ 1997, A\&A, 327, 890  
\reference{}  Richards, G. T., Strauss, M. A., Fan, X., \etal\ 2006, \aj, 131, 2766 
\reference{}  Rudie, G. C., Steidel, C. C., Shapley, A. E., \& Pettini, M.  2013, \apj, 769, 146
\reference{}  Schmidt, T. M.,  Hennawi, J. F., Davies, F. B., \etal\ 2018, \apj, 861, 122 
\reference{}  Shapley, A. E., Steidel, C. C., Strom, A. L.,  \etal\ 2016, \apj, 826, L24 
\reference{}  Shull, J. M., Danforth, C. W., Tilton, E. M., \etal\ 2017, \apj, 849, 106 
\reference{}  Shull, J. M., France, K., Danforth, C. W., \etal\ 2010, \apj, 722, 1312 
\reference{}  Shull, J. M., Roberts, D., Giroux, M. A., Penton, S. V., \& Fardal, M. A. 1999, \aj, 118, 1450 
\reference{}  Shull, J. M., Tumlinson, J., Giroux, M. L., \etal\ 2004, ApJ, 600, 570 
\reference{}  Stevans, M. L, Shull, J. M., Danforth, C. W., \& Tilton, E. M.  2014,  \apj, 794,  75 
\reference{}  Syphers, D., \& Shull, J. M. 2013, ApJ, 765, 119  
\reference{}  Syphers, D., \& Shull, J. M. 2014, ApJ,  784, 42 
\reference{}  Telfer, R., Zheng, W., Kriss, G. A., \& Davidsen, A. F. 2002, \apj, 656, 773 
\reference{}  Tilton, E. M., Stevans, M. L., Shull, J. M., \& Danforth, C. W. 2016, \apj,  817, 56 
\reference{}  Vanzella, E., Nonino, M., Cupani, G., \etal\ 2018, \mnras, 476, L15 
\reference{}  Worseck, G., Prochaska, J. X.,  Hennawi, J. F., \& McQuinn, M. 2016, \apj, 825, 144 
\reference{}  Zheng, W., Kriss, G. A., Telfer, R. \etal\ 1997, \apj, 475, 469 
\reference{}  Zheng, W., Kriss, G. A., Deharveng, J.-M., \etal\ 2004, \apj, 605, 631 

\end{references}
\end{document}